# Threshold current for switching of a perpendicular magnetic layer induced by spin Hall effect

Ki-Seung Lee[1,2], Seo-Won Lee[1], Byoung-Chul Min[2], and Kyung-Jin Lee[1,2,3a)]

[1]*Department of Materials Science and Engineering, Korea University, Seoul 136-701, Korea*

[2]*Spin Convergence Research Center, Korea Institute of Science and Technology, Seoul 136-791, Korea*

[3]*KU-KIST Graduate School of Converging Science and Technology, Korea University, Seoul 136-713, Korea*

**We theoretically investigate the switching of a perpendicular magnetic layer by in-plane charge current due to the spin Hall effect. We find that, in the high damping regime, the threshold switching current is independent of the damping constant, and is almost linearly proportional to both effective perpendicular magnetic anisotropy field and external in-plane field applied along the current direction. We obtain an analytic expression of the threshold current, in excellent agreement with numerical results. Based on the expression, we find that magnetization switching induced by the spin Hall effect can be practically useful when it is combined with voltage-controlled anisotropy change.**

---
a) Electronic mail: kj_lee@korea.ac.kr.

The spin transfer torque [1,2] can cause current-induced magnetization switching in magnetic tunnel junctions (MTJs) consisting of an insulator sandwiched by two ferromagnetic layers; a freely switchable layer (= free layer) and a fixed layer. An electric current running perpendicular to the layers is spin-polarized by the fixed layer and transfers its spin angular momentum to the free layer. This spin-angular-momentum-transfer can exert a spin torque large enough to switch the free layer magnetization. Lots of studies on this subject have addressed its fundamental physics [3-12], and explored its potential applicability to magnetic random access memories (MRAMs) [13-17]. Up to now, most studies have focused on the current-perpendicular-to-plane (CPP) geometry described above.

Several experiments have shown recently that it is also possible to switch the magnetization in current-in-plane (CIP) geometry [18-20]. Liu *et al*. [19,20] demonstrated that an in-plane current flowing in a heavy metal layer attached to a free layer can selectively switch the free-layer magnetization, and reported that these results can be quantitatively explained by spin torque from the spin Hall effect. In ferromagnet|nonmagnet bi-layer systems, an in-plane charge current passing through the nonmagnet is converted into a perpendicular spin current due to the spin Hall effect [21]. The ratio of spin current to charge current is parameterized by spin Hall angle. This spin current, injected perpendicularly to the free layer, transfers its spin-angular-momentum and exerts a spin torque to the free layer magnetization as in the CPP geometry.

Two magnetic configurations have been tested for the magnetization switching induced by the spin Hall effect. One is the in-plane free layer configuration where the free layer has in-plane magnetic anisotropy [19], and the other is the perpendicular free layer configuration where the free layer has perpendicular magnetic anisotropy [18, 20]. For the in-plane free layer configuration, the in-plane charge current running in *x* direction yields a spin current in *z* direction due to the spin Hall effect (see Fig. 1(a) for the coordinate system). The spins injected into a ferromagnet (= free layer) are aligned in ±*y* direction, generating the damping or anti-damping torque when the magnetic easy axis of the free layer is in *y* direction. In this CIP configuration, a threshold current for the free layer switching has the same form as that of the conventional CPP geometry with in-plane free and fixed layers, given as [3, 4, 20]

$$J_{C,\text{in-plane}}^{\text{SH}} = \alpha \frac{2e}{\hbar} \frac{M_S t_F}{\theta_{\text{SH}}} \left( H_{K,\text{in}} + \frac{N_d M_S}{2} \right), \tag{1}$$

where $\alpha$ is the damping constant, $M_S$ is the saturation magnetization, $t_F$ is the thickness of free layer, $\theta_{SH}$ is an effective spin Hall angle of the system, $H_{K,\text{in}}$ is the in-plane magnetic anisotropy field, and $N_d$ is the demagnetization factor, which depends on the patterned shape of free layer.

For the perpendicular free layer configuration, in addition to an in-plane current, an external in-plane field $H_x$ along the direction of current-flow should be applied for deterministic switching [18, 20]. Although this in-plane field does not favor either perpendicular magnetic orientation by itself, it breaks the symmetry in the response to the spin torque and provides the deterministic switching [20]. In this case, however, an explicit expression of the threshold switching current like Eq. (1) has not been reported yet. To design and interpret experiments based on the spin Hall spin torque, it is of critical importance to find such an analytic expression.

In this Letter, we show an analytic expression of threshold switching current for the perpendicular configuration in the CIP geometry. The expression is verified by comparing with numerical results obtained from macrospin simulations. We study the switching of a perpendicular nanomaget on top of a heavy metal layer with flowing in-plane current (the left panel of Fig. 1(a)). To get an insight into the perpendicular switching induced by the spin Hall effect, we numerically solve the modified Landau-Liftshitz-Gilbert equation as [20]

$$\frac{\partial \mathbf{m}}{\partial t} = -\gamma \mathbf{m} \times \mathbf{H}_{\text{eff}} + \alpha \mathbf{m} \times \frac{\partial \mathbf{m}}{\partial t} + \gamma c_J \mathbf{m} \times (\mathbf{m} \times \hat{\mathbf{y}}), \qquad (2)$$

where $\gamma$ (= $1.76 \times 10^7$ Oe$^{-1}$s$^{-1}$) is the gyromagnetic ratio, $\mathbf{H}_{\text{eff}}$ is an effective magnetic field including an effective perpendicular anisotropy field $H_{K,\text{eff}}$ (= $H_K - N_d M_S$) and an external in-plane field $H_x$, $H_K$ is the perpendicular anisotropy field, $c_J$ is $(\hbar/2e)(\theta_{SH} J / M_S t_F)$, $\theta_{SH}$ (= 0.3 [22]) is an effective spin Hall angle of the system, $M_S$ (= 1000 emu/cm$^3$) is the saturation magnetization, and $J$ is the current density in heavy metal layer. The width of heavy metal line ($W$) is set to be the same as the size of nanomagnet ($D$), and the thickness of heavy metal line ($t_N$) is 2 nm. The rise time of the current is 0.5 ns. We conducted macrospin simulations with variables of $\alpha$, $H_{K,\text{eff}}$, $H_x$, $D$, and $t_F$ at zero temperature.

The switching process in this CIP geometry is completely different from the conventional CPP geometry. Figure 1(b) shows the temporal change of magnetization components at a switching current in the CIP geometry. For comparison, a case of conventional CPP geometry with perpendicular free and fixed layers is shown in Fig. 1(c). For each case, a current corresponding to a threshold switching current is applied. An interesting difference can be found in the switching process. The switching occurs via many precessions in the CPP geometry as well-known (Fig. 1(c)), whereas almost no precession is observed for the switching in the CIP geometry (Fig. 1(b)). In the CPP geometry, the spin torque is directed along $\mathbf{m} \times (\mathbf{m} \times \hat{\mathbf{z}})$ and competes with the damping torque directly (i.e., damping torque ≈ $\alpha H_{K,\text{eff}} \mathbf{m} \times (\mathbf{m} \times \hat{\mathbf{z}})$). Since these two torques are almost canceled out at the threshold switching current, magnetization dynamics is dominated by the precession torque, $-\gamma \mathbf{m} \times \mathbf{H}_{\text{eff}}$, resulting in many precessions. This switching process explains why the switching current is proportional to the damping constant in the CPP geometry. In contrast, the spin Hall spin torque in the CIP geometry is directed along $\mathbf{m} \times (\mathbf{m} \times \hat{\mathbf{y}})$ and thus does not compete with the damping torque directly. A similar situation occurs for a layer structure consisting of a perpendicular fixed layer and an in-plane free layer with the current running perpendicular to the plane [23, 24]. In this case, the threshold current is independent of the damping constant and proportional to the anisotropy field of free layer. Consequently, one may expect that the switching current for the CIP geometry studied in this work is also independent of the damping constant.

To check this, we conducted simulations of magnetization dynamics with a current pulse (the pulse width of 5 ns and the rise/fall time of 0.5 ns), varying the damping constant. Fig. 2(a) shows the switching current as a function of the damping constant. For high damping cases ($\alpha > 0.03$), the switching current $I_{SW}$ is independent of $\alpha$ as expected. However, for low damping cases ($\alpha < 0.03$), the switching current randomly jumps between two levels.

Figures 2(b) and 2(c) show time-dependent change in magnetization components at the same current ($I = 135$ μA) for $\alpha = 0.028$ and $\alpha = 0.03$, respectively. In both cases, a significant precessional motion is observed when the current is turned off at $t = 5$ ns ($\equiv$ pulse width). It is because the magnetization direction at $t = 5$ ns significantly deviates from its equilibrium direction, $m_z \approx \pm 1$ (i.e., $|m_y| \gg |m_z|$). For $\alpha = 0.03$ (Fig. 2(c)), $m_z$ eventually goes to −1 and thus switching occurs. Note that $m_z$ is slightly negative at $t = 5$ ns and thus the magnetization is on the downhill towards the point of $m_z = -1$ in the energy landscape. For high damping cases, the magnetization moves towards an energy minimum ($m_z \approx -1$) due to high energy dissipation rate. However, the energy dissipation rate for low damping cases is relatively small, giving rise to a non-linear dynamics. For example, when $\alpha = 0.028$ (Fig. 2(b)), $m_z$ switches back to +1. This switching-back may be understood by that a highly nonlinear precession dynamics disturbs the complete switching. The reason of two level fluctuations of the switching current in the low damping regime (Fig. 2(a)) is not clearly understood but may be related to the period-doubling bifurcation of chaos theory.

The results shown in Fig. 2 suggest that a high damping is required for practical applications of CIP switching. We note that the damping constant of perpendicular magnetic materials is usually large, and the heavy metal in contact with the nanomagnet can further increase the damping due to spin pumping effect [25] or spin motive force [26]. For instance, Mizukami *et al.* [27] reported experimental results that the damping constant of Pt|Co|Pt structure with perpendicular anisotropy is larger than 0.1. Based on this, we assume a high damping ($\alpha$ = 0.1) in the remaining part of this paper.

The damping-independent switching current in high damping regime indicates that the threshold switching current can be obtained from a static solution of Eq. (2); i.e. $\partial \mathbf{m}/\partial t = 0$. Then, the resultant equation with $\mathbf{m} = (\cos\phi\sin\theta, \sin\phi\sin\theta, \cos\theta)$ is

$$H_x \cos\theta - H_{K,\text{eff}} \cos\phi \cos\theta \sin\theta + c_J (\cos^2\theta + \cos^2\phi \sin^2\theta) = 0. \qquad (3)$$

Another important implication of the modelling result shown in Fig. 2(c) is that $m_y$ component is almost zero before the switching of $m_z$ component, as indicated by a down-arrow in Fig. 2(c). Then, the switching of $m_z$ component to a negative value is accompanied by an abrupt change in $m_y$ component as indicated by an up-arrow. This suggests that the threshold switching current can be obtained by the stability condition of Eq. (3) with $\phi$ = 0. In other words, the threshold $c_J$ is determined by the condition that there is no $\theta$ satisfying $H_x \cos\theta - H_{K,\text{eff}} \cos\theta \sin\theta + c_J = 0$. From this stability condition, one finds the threshold $c_J$ (= $c_{SW}$)

$$c_{SW} = \sqrt{\frac{H_{K,\text{eff}}^2}{32}\left[8 + 20\left(\frac{H_x}{H_{K,\text{eff}}}\right)^2 - \left(\frac{H_x}{H_{K,\text{eff}}}\right)^4 - \left(\frac{H_x}{H_{K,\text{eff}}}\right)\left(8 + \left(\frac{H_x}{H_{K,\text{eff}}}\right)^2\right)^{3/2}\right]}. \qquad (4)$$

Using Eq. (4) and $c_J$ = $(\hbar/2e)(\theta_{SH} J / M_S t_F)$, the switching current density is given by

$$J_{C,\text{perp}}^{\text{SH}} = \frac{2e}{\hbar} \frac{M_S t_F}{\theta_{\text{SH}}} \left( \sqrt{\frac{H_{K,\text{eff}}^2}{32} \left[ 8 + 20 \left( \frac{H_x}{H_{K,\text{eff}}} \right)^2 - \left( \frac{H_x}{H_{K,\text{eff}}} \right)^4 - \left( \frac{H_x}{H_{K,\text{eff}}} \right) \left( 8 + \left( \frac{H_x}{H_{K,\text{eff}}} \right)^2 \right)^{3/2} \right]} \right).$$

(5)

When $H_x \ll H_{K,\text{eff}}$, Eq. (5) is further simplified as

$$J_{C,\text{perp}}^{\text{SH}} = \frac{2e}{\hbar} \frac{M_S t_F}{\theta_{\text{SH}}} \left( \frac{H_{K,\text{eff}}}{2} - \frac{H_x}{\sqrt{2}} \right).$$

(6)

Equations (5) and (6) are the central results of this work. To verify the validity of these analytic expressions, we numerically obtain the threshold switching currents in wide ranges of $H_{K,\text{eff}}$ (0.25 T to 1.2 T) and $H_x$ (50 Oe to 1000 Oe) for a nanomagnet with $D = 30$ nm and $t_F = 1$ nm. Fig. 3(a) and 3(b) show numerically obtained switching current as a function of $H_{K,\text{eff}}$ and $H_x$, respectively. We find the threshold switching current is almost linearly proportional to both $H_{K,\text{eff}}$ and $H_x$ as expected from Eq. (6). Figure 3(c) shows the errors of Eqs. (5) and (6), defined by |numerical $J_{C,\text{perp}}^{\text{SH}}$ − analytic $J_{C,\text{perp}}^{\text{SH}}$|/(numerical $J_{C,\text{perp}}^{\text{SH}}$) × 100 %, as a function of the thermal stability factor $\Delta_{\text{SH}}$ (= $E_B/k_B T$) at room temperature. Here, the energy barrier $E_B$ is $E_B = H_{K,\text{eff}} M_S V \left( 1 - (H_x / H_{K,\text{eff}})^2 \right) / 2$. We find that when $\Delta_{\text{SH}} > 40$, the errors are less than 2 % for Eq. (5) and 4 % for Eq. (6), respectively. These errors may be comparable to or even less than the inaccuracy in experiments. We also test the applicability of Eqs. (5) and (6) for various $D$ (10 nm to 30 nm) and $t_F$ (1 nm to 3 nm), and find that the analytic expressions can describe the threshold switching current density with a sufficiently good accuracy as for an example shown in Fig. 3(c). Therefore, Eqs. (5) or (6) can be used to design and interpret experiments performed for current-induced perpendicular switching of magnetization induced by the spin Hall effect.

We note that Eq. (5) is applicable to any ferromagnet|non-magnet bi-layer structures

when the non-magnetic layer is able to supply a sufficient spin Hall spin current. The nonmagnetic layer could be either a heavy metal such as Pt, Ta, and W [19, 20, 22] or an alloy consisting of light host materials and heavy metal impurities such as CuIr [28] and CuBi [29]. The magnitude and sign of the spin Hall angle as well as the damping constant of the bi-layer structure strongly depend on the properties of nonmagnetic layer. For a non-magnetic layer with the opposite sign of spin Hall angle, one finds exactly the same symmetry of the torques by reversing the direction of $H_x$. Concerning the high damping condition, the authors of Ref. [20] reported that there are two switching boundaries depending on $H_x$; one occurs for a small $H_x$ (= type I), and the other occurs for a large $H_x$ (= type II). We found that the high damping condition for the controllable switching described in Fig. 2(a) is required only for the type I. Therefore, if a bi-layer structure has a low damping constant, one can simply increase $H_x$ to avoid uncontrollable switching. The different switching boundary for different type of switching in Ref. [20] is caused by a very slow ramping rate of the current [30]. In our case, we assumed a fast ramping rate (i.e., current-rise time = 0.5 ns), which is a reasonable assumption for realistic device applications. We found that in this fast ramping condition, the switching current of both types of switching is successfully described by Eq. (5).

We next discuss about the usefulness of perpendicular switching induced by the spin Hall spin torque. For the conventional CPP geometry with perpendicular free and fixed layers, the switching current density $J_{\text{C,perp}}^{\text{CONV}}$ is given by $J_{\text{C,perp}}^{\text{CONV}} = \alpha(2e/\hbar)(M_S t_F/\eta)H_{\text{K,eff}}$ [15]. Using Eq. (6) and assuming that $H_x/H_{\text{K,eff}}$ is very small, we find $J_{\text{C,perp}}^{\text{SH}}/J_{\text{C,perp}}^{\text{CONV}} \approx (1/2\alpha)(\eta/\theta_{\text{SH}})$. This ratio is $0.55/\alpha$ for parameters of $\eta = 0.33$ [8], and $\theta_{\text{SH}} = 0.3$ [22]. For typical ranges of $\alpha$ of perpendicular materials (0.03 [16] to 0.1 [27]), the ratio $J_{\text{C,perp}}^{\text{SH}}/J_{\text{C,perp}}^{\text{CONV}}$ is 5.5 to 18, which is not small. The absolute value of $J_{C,perp}^{SH}$ is indeed large, of the order of $10^8$ A/cm$^2$ for $\Delta_{SH} >$ 40, indicating that magnetization switching induced by the spin Hall effect alone may not be practically attractive for application. This disadvantage may be overcome by voltage-induced control of the free-layer magnetic anisotropy; a series of recent experiments have shown that the magnetic anisotropy is controlled by a voltage application across MTJs [31-35]. In the conventional CPP geometry, $H_{K,eff}$ can be either increased or decreased depending on the voltage polarity during spin-transfer switching, whereas, in the CIP geometry with the spin Hall effect, $H_{K,eff}$ can be readily decreased by applying a unipolar voltage bias across a thick insulating barrier regardless of the in-plane switching current direction. It was demonstrated that for the in-plane free-layer magnetization [34, 35], this voltage-controlled anisotropy change can be combined with the spin Hall effect to reduce the switching current density. The same approach can be applied for the perpendicular free-layer magnetization. In this case, $H_{K,eff}$ in Eqs. (5) and (6) is replaced by $H_{K,eff} - V_{app}[d(H_{K,eff})/dV_{app}]$, where $V_{app}$ is the applied voltage across MTJs and $d(H_{K,eff})/dV_{app}$ describes the voltage-induced change in the magnetic anisotropy. Liu *et al.* found $d(H_{K,eff})/dV_{app} \approx$ 730 Oe/V in Ta|CoFeB|MgO|CoFeB MTJs [35]. Figure 3(d) shows that $J_{C,perp}^{SH}$ decreases rapidly with increasing $d(H_{K,eff})/dV_{app}$ and $J_{C,perp}^{SH}$ can be reduced below $10^7$ A/cm$^2$ for $d(H_{K,eff})/dV_{app} >$ 3.3 kOe/V. Our result suggests that the magnetization switching induced by a combined effect of spin Hall spin torque and voltage-controlled anisotropy change could be practically useful with an enhanced $d(H_{K,eff})/dV_{app}$.

Finally, we remark that in contrast to the spin Hall effect, Miron *et al.* proposed that Rashba-type spin-orbit coupling is mainly responsible for spin torque resulting in the perpendicular switching induced by an in-plane current [18]. The existence of this Rashba effect in metallic ferromagnets is a subject under extensive discussion [36-42]. Although we do not consider the Rashba effect in this work, a further investigation about this possibility may be valuable.

To summarize, we investigate the threshold switching current density for in-plane current-induced perpendicular switching of magnetization due to the spin Hall effect. We find that the switching current in high damping regime is independent of the damping constant, and is in almost linear relation with both effective perpendicular anisotropy field and external magnetic field applied along the current direction. We derive an explicit analytic expression of threshold switching current and verify its applicability by testing various cases numerically. This expression will be of importance for both fundamental physics and applications since it can be used to estimate essential physical quantities such as spin Hall angle and to design practical devices utilizing the spin Hall effect.

This work was supported by the NRF (2010-0023798, 2011-0028163), by the MEST Pioneer Research Center Program (2011-0027905), and KU-KIST School Joint Research Program.

**FIGURE CAPTION**

FIG. 1. (Color online) (a) Schematic diagrams of model system: (left) Current-In-Plane (CIP) geometry utilizing spin transfer torque from the spin Hall effect, (right) Current-Perpendicular-to-Plane (CPP) geometry utilizing conventional spin transfer torque. Time-dependent change in magnetization components for (b) in-plane current-induced perpendicular switching due to the spin Hall effect (= CIP geometry), and (c) conventional current-induced perpendicular switching (= CPP geometry). For (b) and (c), following modeling parameters are used; $\alpha = 0.1$, $H_{K,eff} = 0.4$ T, $H_x = 200$ Oe, $D = 30$ nm, and $t_F = 1$ nm.

FIG. 2. (Color online) (a) Switching current as a function of the damping constant $\alpha$. Time-dependent change in magnetization components for (b) $\alpha = 0.028$, and (c) $\alpha = 0.03$. Following modeling parameters are used; $H_{K,eff} = 0.469$ T, $H_x = 200$ Oe, $D = 30$ nm, and $t_F = 1$ nm. Top panels in (b) and (c) show the current-pulse profile (pulse width = 5 ns and rise/fall time = 0.5 ns).

FIG. 3. (Color online) Numerical results of the switching current as a function of (a) $H_{K,eff}$, and (b) $H_x$. (c) Errors of the analytical expressions (Eqs. (5) and (6)) of threshold switching current density as a function of the thermal stability factor $\Delta_{SH}$. The errors and $\Delta_{SH}$ are defined in the main text. (d) Switching current density as a function of $d(H_{K,eff})/dV_{app}$. The following parameters are used for (d): $\Delta_{SH} = 40$, $H_x = 1000$ Oe, $\theta_{SH} = 0.3$, $D = 30$ nm, and $t_F = 1$ nm.

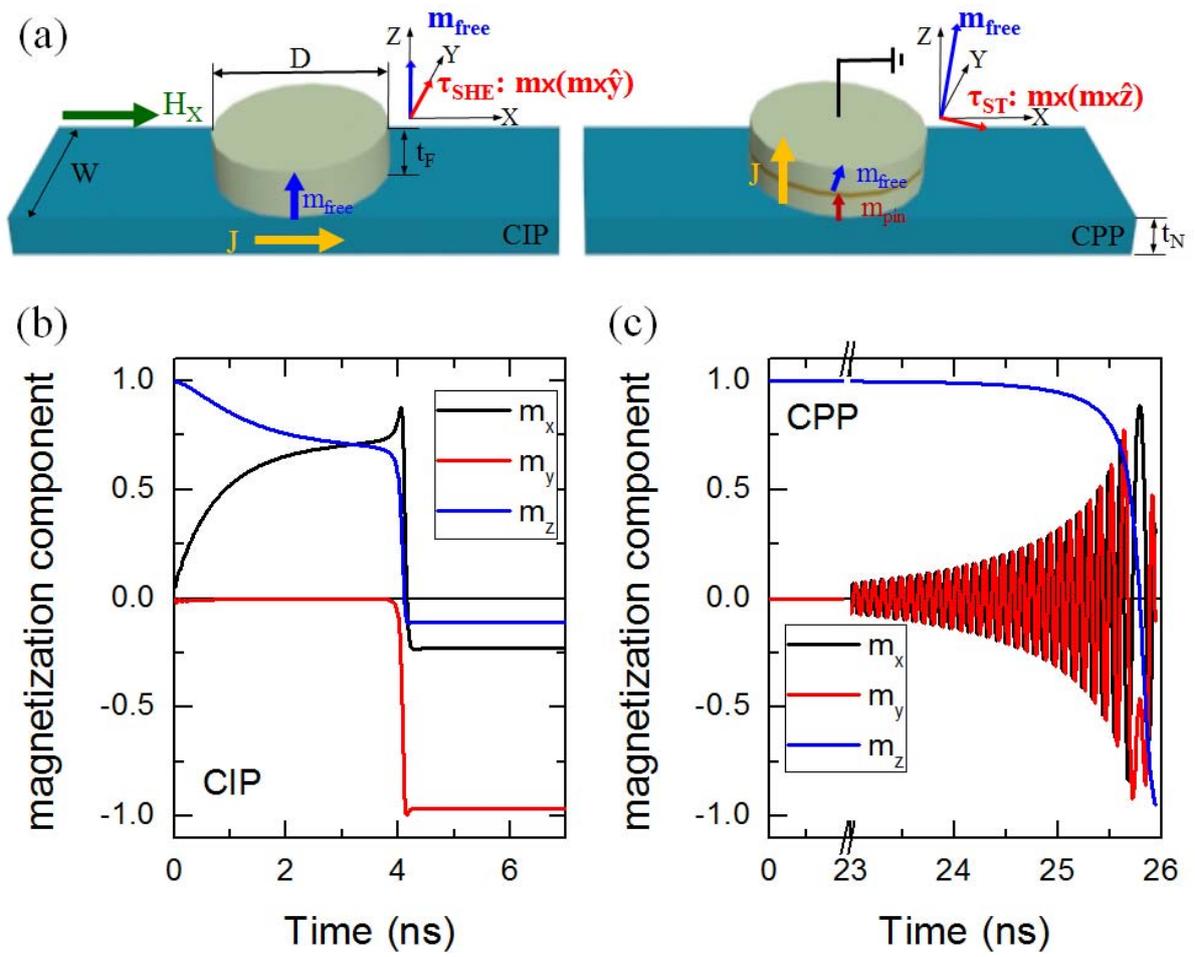

FIG. 1. Lee *et al*.

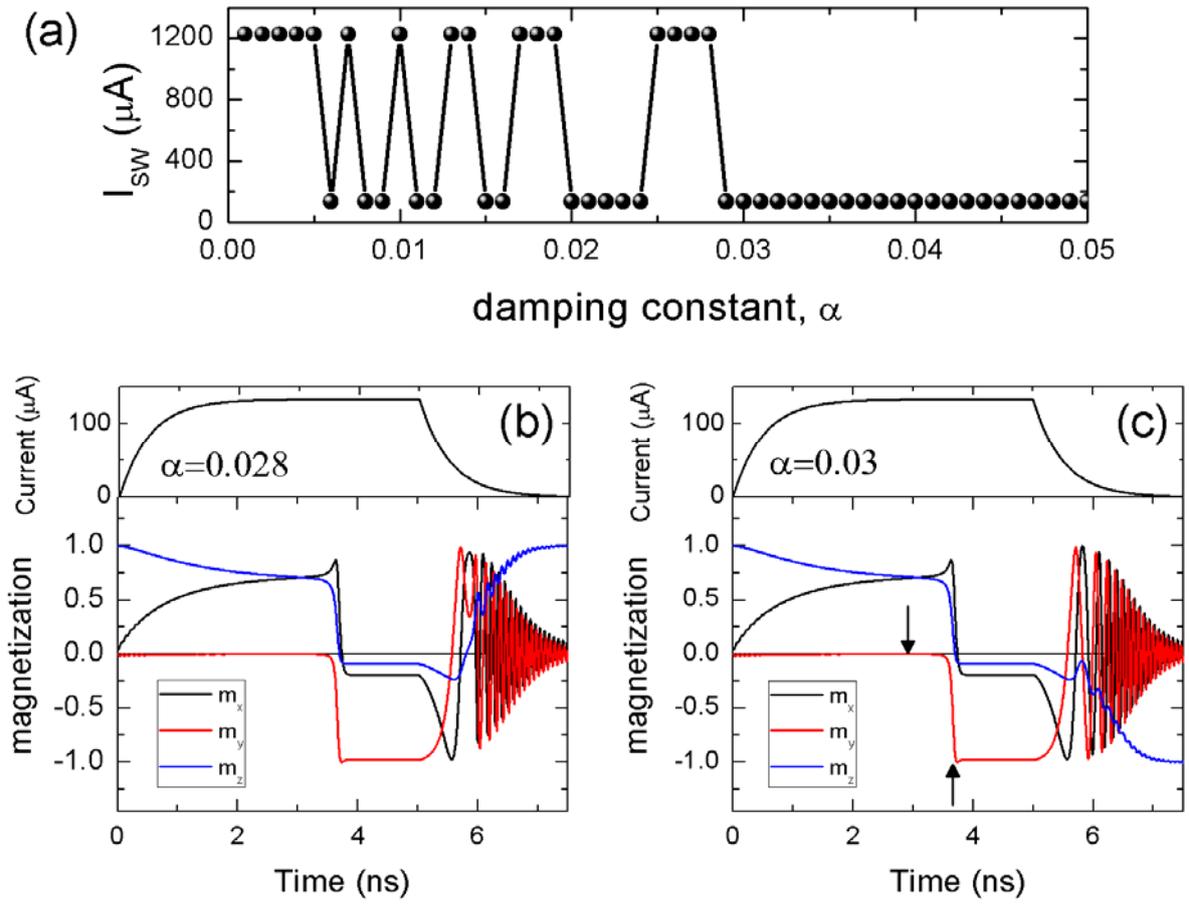

FIG. 2. Lee *et al*.

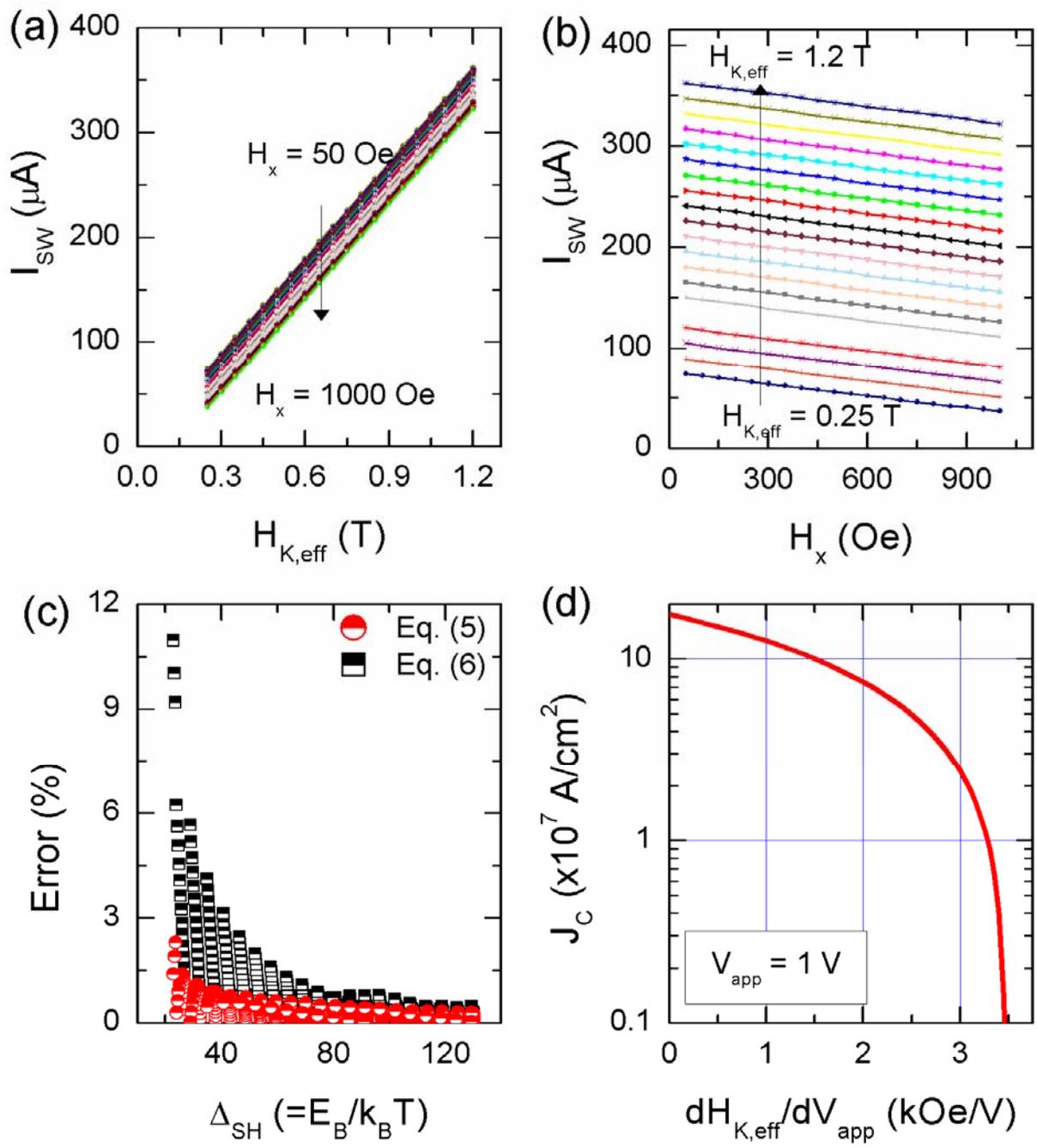

FIG. 3. Lee *et al*.